%
%
%

\ifx\mnmacrosloaded\undefined \input mn\fi

%

\newif\ifAMStwofonts

\ifCUPmtplainloaded \else
  \NewTextAlphabet{textbfit} {cmbxti10} {}
  \NewTextAlphabet{textbfss} {cmssbx10} {}
  \NewMathAlphabet{mathbfit} {cmbxti10} {} 
  \NewMathAlphabet{mathbfss} {cmssbx10} {} 
  \ifAMStwofonts
    \NewSymbolFont{upmath} {eurm10}
    \NewSymbolFont{AMSa} {msam10}
    \NewMathSymbol{\upi}     {0}{upmath}{19}
    \NewMathSymbol{\umu}     {0}{upmath}{16}
    \NewMathSymbol{\upartial}{0}{upmath}{40}
    \NewMathSymbol{\leqslant}{3}{AMSa}{36}
    \NewMathSymbol{\geqslant}{3}{AMSa}{3E}

  \else
    \def\umu{\mu}
    \def\upi{\pi}
    \def\upartial{\partial}
  \fi
\fi


\pageoffset{-2.5pc}{0pc}

\loadboldmathnames



\pagerange{000--000}    
\pubyear{0000}
\volume{000}

\begintopmatter  

\title{Reclassification of gamma-ray bursts}

\author{Andreu Balastegui$^{{\rm 1}}$, 
Pilar Ruiz-Lapuente$^{{\rm 1}{\rm ,}{\rm 2}}$ and 
Ramon Canal$^{{\rm 1}{\rm ,}{\rm 3}}$}

\affiliation{$^{{\rm 1}}$Departament d'Astronomia i Meteorologia, Universitat 
de Barcelona, Mart\'{\i} i Franqu\'{e}s 1, Barcelona 08028, Spain \break
$^{{\rm 2}}$Max-Planck-Institut f\"{u}r Astrophysik, 
Karl-Schwarzschild-Strasse 1, 85740 Garching bei M\"{u}nchen, 
Germany \break $^{{\rm 3}}$Institut d'Estudis Espacials de Catalunya, Nexus 
Building, 2-4 Gran Capit\`{a}, Barcelona 08034, Spain}

\shortauthor{Balastegui et al.}
\shorttitle{Reclassification of gamma-ray bursts}

\acceptedline{Accepted 0000 . Received 0000 ; in original form 0000}

\abstract {We have applied two different automatic classifier algorithms to 
the BATSE Current GRB Catalog data and we obtain three different classes of 
GRBs. Our results confirm the existence of a third, intermediate class of 
GRBs, with mean duration $\sim$ 25-50 s, as deduced from a cluster analysis 
and from a neural network algorithm. Our analyses imply longer durations than 
those found by Mukherjee et al. (1998) and Horv\'{a}th (1998), whose  
intermediate class had durations $\sim$ 2-10 s. From the neural network 
analysis no difference in hardness between the two longest classes is found, 
and from both methods we find that the intermediate-duration class 
constitutes the most homogeneous sample of GRBs in its space distribution 
while the longest-duration class constitutes the most inhomogeneous 
one with $<$V/V$_{{\rm m}{\rm a}{\rm x}}$$>$ $\sim$ 0.1, being thus the 
deepest population of GRBs with z$_{{\rm m}{\rm a}{\rm x}} \sim$ 10. The 
trend previously found in long bursts, of spatial inhomogeneity increasing 
with hardness, only holds for this new longest-duration class.}


\maketitle 

\note{{}}{{$^{{\rm 1}}$E-mail: abalaste@am.ub.es}}
\note{{}}{{$^{{\rm 2}}$E-mail: pilar@am.ub.es}}
\note{{}}{{$^{{\rm 3}}$E-mail: ramon@am.ub.es}}

\section{Introduction}

Since their discovery in the late 60's (Klebesadel, Strong \& Olson 
1973), GRBs have been a long-remaining puzzle (see Piran 2000; 
M\'{e}sz\'{a}ros 2001; Castro-Tirado 2001, for recent reviews). Models 
involving a short distance scale implied an emitted energy 
$\sim$10$^{{\rm 4}{\rm 2}}$ erg, whereas a cosmological origin required 
10$^{{\rm 5}{\rm 1}}$ erg at least. The increasing degree of isotropy found 
as the sample of GRBs grew, plus the lack of faint GRBs (Meegan, Fishman \& 
Wilson 1985) favored a cosmological scenario. Moreover, after the launch, in 
1991, of the BATSE instrument on board the {\it CGRO}, a very high degree of 
isotropy was found in the new, much larger sample. Finally, the measurement 
in 1997 of the first GRB redshift (Metzger et al. 1997) and the subsequent 
ones have confirmed that GRBs are at cosmological distances.

Concerning the physical mechanism of GRBs, the fireball shock model (Rees \& 
M\'{e}sz\'{a}ros 1992; M\'{e}sz\'{a}ros \& Rees 1993; Daigne \& Mochkovitch 
1998) is a progenitor-independent model for radiation emission that succeeds 
in explaining both the burst itself and its afterglow. There is, however, a 
variety of proposed objects that are capable of generating the GRBs (Nemiroff 
1994): from mergings of neutron stars with neutron stars (Paczy{\'n}ski 1990) 
or with black holes (Narayan et al. 1992) to collapsars (Woosley 1993; 
MacFadyen \& Woosley 1999) and hypernovae (Paczy{\'n}ski 1998). Magnetic 
instability in a neutron star being spun up by accretion in an X-ray binary 
could also produce them (Spruit 1999). There are more exotic models, 
involving quark stars (Ma \& Xie 1996), mirror stars (Blinnikov 1999), or 
cosmic strings (Berezinsky, Hnatyk \& Vilenkin 2001). In the face of such a 
boiling pot of theoretical ideas, to know the positions of the GRBs with 
respect to their host galaxies and specially their redshifts are key issues 
(Bloom et al. 2000). Once a fair sample of redshifts becomes available, which 
should happen soon with the up-to-date technology of new missions like HETE 
II and {\it Swift}, the distances will be known and with that the released 
energy and the luminosity function of GRBs, together with their distribution 
across the Universe. That should certainly discriminate among existing 
models, and it should give as well unprecedented information on the very 
structure of the Universe up to redshifts far higher than 5 (Lamb \& Reichart 
2000; hereafter LR00), on the cosmic star formation history (Totani 1999), 
and on the evolution of galaxies (Totani 1997).

It is likely that more than strictly one progenitor could give rise to GRBs, 
since it has been shown that different objects can produce a burst 
of gamma rays with the observed characteristics. Therefore, the catalog of 
GRBs may reflect the manifestations of various phenomena. To uncover them,  
various attempts to separate different classes of GRBs have been made.

The two most generally accepted classes of GRBs are those arising from the 
bimodal distribution of their durations (Kouveliotou et al. 1993; hereafter 
K93), that separates long (lasting for more than 2 s) and short 
(less than 2 s) GRBs, the short bursts being at the same time spectrally 
harder than the long bursts. The different spatial distribution of the two 
classes has also been shown (Katz \& Canel 1996), and it is consistent with 
isotropy in both. However, the longer GRBs appear to be more inhomogeneous  
in their space distribution than the shorter ones, as deduced from the higher 
value of $<$V/V$_{{\rm m}{\rm a}{\rm x}}$$>$ for the short GRBs (this quantity 
measures the deviation of the space distribution from a homogeneous Euclidean
distribution). Separating the long bursts class in two groups of hardness 
H$_{{\rm 3}{\rm 2}}$ (the fluence ratio of spectral channel 3 to spectral 
channel 2), respectively higher and lower than 3, Tavani (1998; hereafter 
T98) found that the long/hard bursts are more inhomogeneously distributed 
than the long/soft ones. We will show here that there exists, indeed, a 
trend, among the long bursts, of harder bursts being more inhomogeneous in
their space distribution, which holds for hardnesses up to $\sim$4, with a 
slight rise of $<$V/V$_{{\rm m}{\rm a}{\rm x}}$$>$ beyond that point. In the 
three-class classification that we propose here this trend, however, only 
exists for the longest class.

In the following we first describe the two methods (cluster analysis and 
neural network) that are used to classify GRBs in the present work. The 
results from the two different methods are then discussed and compared. We 
find that both methods point to a classification in three classes, that 
resulting from splitting the 'old' long-burst class in two, and we compare 
the characteristics of the new three groups. Finally, the possible physical 
meaning of the new classification is discussed.

\section{Methodology}

The usual approach in GRB classification has been based on the study of 
bivariate distributions. However, as noted by Bagoly et al. (1998; hereafter 
B98), the BATSE catalog provides up to nine quantities intrinsic to the 
burst (7 related to energy and 2 related to duration), plus other quantities 
corresponding to spatial distribution and to errors in the magnitudes.  
New composite quantities can also be defined, such as the different measures 
of spectral hardness (from the fluence ratios in different spectral channels), 
and also V/V$_{{\rm m}{\rm a}{\rm x}}$. That involves a large number of 
variables which is difficult to handle, complex relationships among them 
(including nonlinear ones) being likely present. Such relationships can 
hardly be directly visualized, and thus multivariate analysis is needed.

Starting from the BATSE Current GRB Catalog (available at 
http://www.batse.msfc.nasa.gov/batse/grb/catalog

\noindent
/current) in its version of September 2000, 1599 bursts have been selected:  
those for which nonzero values of all nine magnitudes are given. These 
magnitudes are the four time-integrated fluences F$_{\rm Ch1}$-F$_{\rm Ch4}$, 
respectively corresponding to the 20-50 keV, 50-100 keV, 100-300 keV, and 
300+ keV spectral channels; the three peak fluxes P$_{\rm 64}$, P$_{\rm 256}$, 
and P$_{\rm 1024}$, measured in 64, 256, and 1024 ms bins, respectively; and 
the two measures of burst duration T$_{{\rm 50}}$ and T$_{{\rm 90}}$, the 
times within which 50\% and 90\% of the flux arrives. Then a principal 
component analysis (PCA) of the standardized logarithms (zero mean and unity 
variance) of those quantities has been performed, obtaining results (Table 1) 
that are very similar to those of B98. As it is well known, PCA is a 
statistical method used in multivariate data analysis to obtain new 
variables, linear combinations of the original ones, which carry most of the 
variance of the system. Based on the correlations among the original 
variables, some of the new variables can be disregarded if they carry very 
little information. For further details on PCA, the reader is referred to B98 
and to Murtagh \& Heck (1987; hereafter M87).

\begintable*{1}
\caption{{\bf Table 1.} Principal Component Analysis for the standardized 
logarithms of fluences, peak fluxes and durations. There is shown, in each 
row, the components of each principal axis in the base of our original 
variables (columns), together with the percentage of the variance carried by 
each of the new axes (first column). For instance, the first principal 
component is: 
$-$ 0.39 log F$_{{\rm Ch 1}}$ $-$ 0.40 log F$_{{\rm Ch 2}}$ $-$ 0.40 log 
F$_{{\rm Ch 3}}$ $-$ 0.33 log F$_{{\rm Ch 4}}$ $-$ 0.22 log P$_{{\rm 64}}$ 
$-$ 0.28 log P$_{{\rm 256}}$ $-$ 0.36 log P$_{{\rm 1024}}$ $-$ 0.29 log 
T$_{{\rm 50}}$ $-$ 0.30 log T$_{{\rm 90}}$.}
\halign{%
\rm\hfil#&\qquad\rm#\hfil&\qquad\rm#\hfil&\qquad\rm#\hfil
&\qquad\rm#\hfil&\qquad\rm#\hfil&\qquad\rm#\hfil
&\qquad\rm#\hfil&\qquad\rm#\hfil&\qquad\rm#\hfil&\qquad\hfil\rm#\cr
\noalign{\hrule}
\noalign{\hrule}
\noalign{\vskip 2pt}
{\bf \%}&{\bf log F$_{{\rm Ch 1}}$}&{\bf log F$_{{\rm Ch 2}}$}&{\bf log 
F$_{{\rm Ch 3}}$}&{\bf log F$_{{\rm Ch 4}}$}&{\bf log P$_{{\rm 64}}$}&{\bf 
log P$_{{\rm 256}}$}&{\bf log P$_{{\rm 1024}}$}&{\bf log T$_{{\rm 50}}$}&{\bf 
log T$_{{\rm 90}}$}\cr
\noalign{\vskip 2pt}
\noalign{\hrule}
\noalign{\hrule}
\noalign{\vskip 3pt}
64.3&$-$0.39&$-$0.40&$-$0.40&$-$0.33&$-$0.22&$-$0.28&$-$0.36&$-$0.29&$-$0.30\cr
27.0&+0.15&+0.12&+0.04&$-$0.05&$-$0.53&$-$0.47&$-$0.30&+0.44&+0.41\cr
4.9&$-$0.22&$-$0.19&+0.06&+0.92&$-$0.10&$-$0.13&$-$0.17&$-$0.06&$-$0.08\cr
1.7&+0.48&+0.41&+0.21&+0.03&$-$0.22&$-$0.25&$-$0.13&$-$0.47&$-$0.46\cr
0.8&+0.56&$-$0.05&$-$0.77&+0.19&+0.16&+0.06&$-$0.09&$-$0.01&+0.12\cr
0.6&+0.01&+0.11&+0.19&$-$0.06&+0.60&$-$0.03&$-$0.75&+0.13&$-$0.00\cr
0.4&$-$0.02&$-$0.05&+0.11&$-$0.02&+0.04&$-$0.04&$-$0.07&$-$0.69&+0.71\cr
0.2&+0.49&$-$0.78&+0.38&$-$0.07&$-$0.04&+0.05&$-$0.03&+0.04&$-$0.06\cr
0.1&$-$0.01&+0.08&+0.01&+0.00&$-$0.46&+0.78&$-$0.40&$-$0.03&+0.01\cr
\noalign{\vskip 3pt}
\noalign{\hrule}
\noalign{\hrule}
}

\endtable

As seen in Table 1, with only three variables, linear combinations of the 
original ones, $\sim$96\% of the system information can be accounted for. The 
first row shows that 64\% of the information is carried by a variable which 
is a weighted sum of of all the original variables with nearly the same weight 
for each of them. The second principal component in importance is 
approximately the difference between the weighted sum of the logarithms of 
the three peak fluxes and that of the logarithms of the two durations, all 
again with similar weights. And with 5\% of the total variance of the system 
the logarithm of the fluence in the fourth channel is found. 

Our current goal is to achieve an automatic classification based on the nine 
original variables, and for that two different methods are used: a cluster 
analysis applied to the results of the PCA, and a neural network algorithm.

\subsection{Cluster Analysis}

For the cluster analysis the {\sc midas} statistical package has been used. 
As stated above, a PCA is first preformed. In this way are obtained new
variables into which the problem becomes easier to separate. This result 
provides the starting point for the cluster analysis, where the Ward's 
criterion of minimum variance (Ward 1963; see also M87) is used. The 
analysis follows an agglomerative hierarchical clustering procedure, which 
starts from n points spread over the 9-dimensional space and groups them 
until ending up with a single cluster. The algorithm searches for clusters 
with minimum variance among objects belonging to the same cluster and with 
maximum variance between clusters, and works with the 'center of gravity' of 
each cluster. That gives clusters as compact and as detached from each other 
as possible.

A dendrogram is obtained, that shows the way groups are clustering, as well 
as the inner variance of the resulting groups. Thus, detecting a large rise 
in the variance by the union of two clusters means that two groups with 
remarkably different characteristics have been merged.

It is important to notice that the PCA looks for combinations of variables 
to obtain axes which have the maximum possible variance. As seen above, 
with just three of those axes more than 95\% of the total variance is 
accounted for. The cluster analysis could, therefore, have been applied to 
those three new variables only. That has been checked and it is found that 
the results are barely affected by this reduction in the number of variables, 
since most of the system information is conserved. Here, however, the results 
of the complete 9-dimensional analysis will be presented, since those same 
nine variables will later be used for the neural network analysis.

Concerning the previous work of Mukherjee et al. (1998; hereafter M98), 
it must be noted that they choose six variables for their analysis, three of 
them being the peak flux log P$_{\rm 256}$ plus the two durations 
(log T$_{\rm 50}$ and log T$_{\rm 90}$), the other three being the total 
fluence log F$_{{\rm total}}$ and two hardnesses, 
log H$_{{\rm 3}{\rm 2}{\rm 1}}$ and log H$_{{\rm 3}{\rm 2}}$. It has been 
learned, from the PCA, that three variables are necessary, which together 
carry more than 95\% of the variance, one of them being approximately the 
logarithm of the fluence in the fourth channel, F$_{\rm Ch4}$ (see above). 
So it seems that M98 do lose information by not taking into account the 
fluence in the fourth channel separately, and also by not considering any 
peak flux later on. It must equally be noticed, when comparing results, that 
our GRB sample is twice as large as theirs.

The main weakness of the cluster analysis is that it only deals with linear 
combinations of the variables. Such a weakness can be overcome by means of a 
neural network analysis, which also detects non-linear relationships.

\subsection{Neural Network}

Neural networks are artificial intelligence algorithms that can be used for 
an automatic and objective classification. We do not want to start from any 
prior classification. Therefore, a  non-supervised algorithm is used. As 
we do not wish to introduce any tracer object either, the net is initialized 
at random. The 'Self-Organizing Map' algorithm (Kohonen 1990), 
implemented in the {\sc som\_pak} package from the Laboratory of Computer and 
Information Science of the University of Helsinki, is used.

As in the cluster analysis, the entrance parameters are the logarithms 
of the same nine variables.

The dimension of the output space must be specified, and based on the results 
of the cluster analysis the network is run two times asking first for a two 
and then for a three-dimensional output space, thus grouping either two or 
three classes of GRBs. The net is trained in two steps before looking for 
results.

\section{Results}

\subsection{Cluster Analysis}

In Fig. 1 the dendrogram with the last six levels of clustering is shown. It
can be seen that the first important increase of the variance occurs when 
joining group 3 with group 2, which tells that two groups with somewhat 
different characteristics have been merged, but the most significant rise in 
variance occurs when merging cluster 2 with cluster 1. From that it is 
concluded that there are two well-separated classes plus an emergent third 
class. 

\beginfigure{1}

\input epsf
\centerline{ \epsfysize=55mm \epsfbox{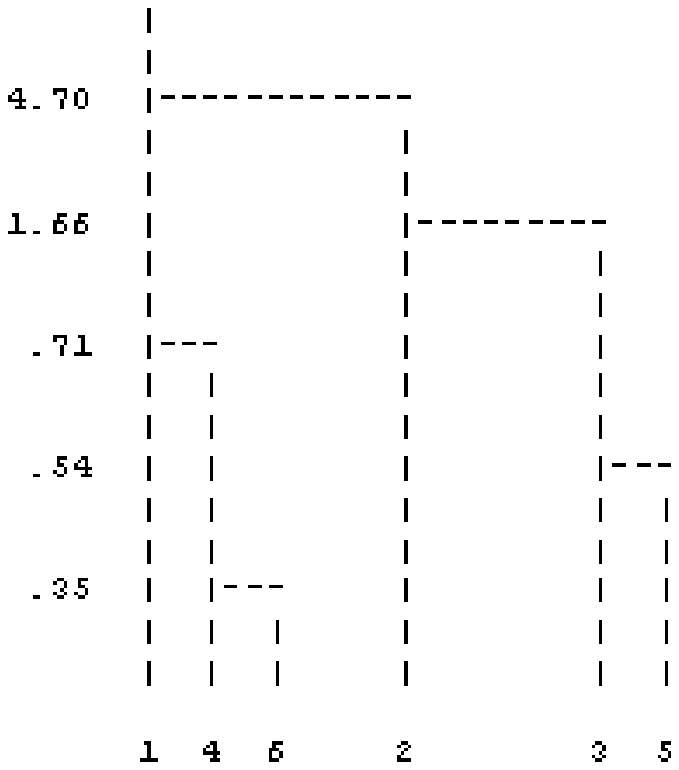} }
\caption{{\bf Figure 1.} Dendrogram of the 9-dimensional analysis. The numbers
at the bottom of the diagram are identifiers of the groups, and those 
at the left show the inner variance of the groups. For instance, when merging 
group 6 with group 4 the variance of the cluster is 0.35. The largest 
increment in the variance occurs when merging group 2 with group 1, with a 
variance increase of 3.04.}

\endfigure

Fig. 2 shows what happens when adding, to the nine starting variables, 
the two extra variables H$_{{\rm 3}{\rm 2}}$ and 
$<$V/V$_{{\rm m}{\rm a}{\rm x}}$$>$. In that case the sample is reduced to 
757 bursts only (instead of 1599), for which all the eleven quantities are 
known.  It can be seen that the three-class classification is the most 
favoured one.

\beginfigure{2}

\input epsf
\epsfysize=55mm \epsfbox{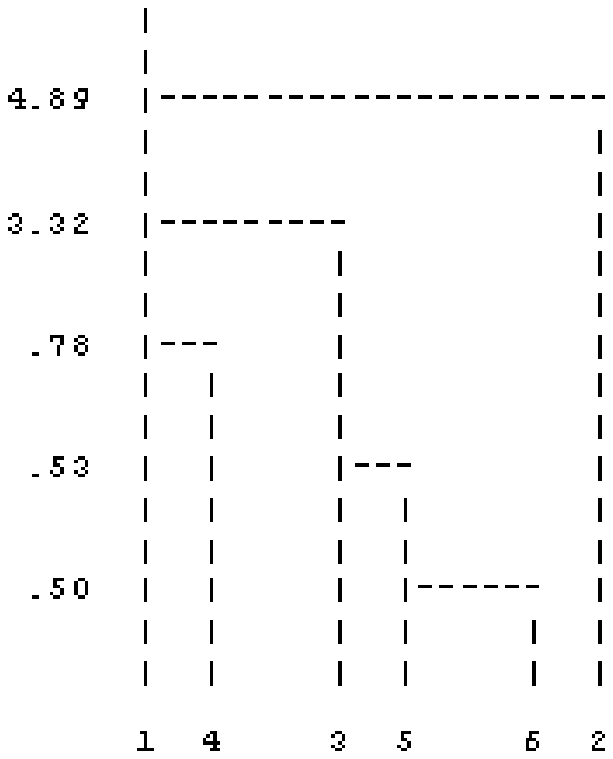}
\caption{{\bf Figure 2.} Dendrogram of the 11-dimensional analysis. Here the 
largest increase in the clusters variance occurs when joining groups 3 and 
1, that suggesting three different classes of GRBs.}

\endfigure

Next, in Table 2, the main characteristics of each GRB class are shown, 2-I 
and 2-II corresponding to the two-class classification, and 3-I, 3-II and 
3-III corresponding to the three-class classification. The deviations 
correspond to $\sigma/\sqrt{N-1}$. The results of the 11-dimensional cluster 
analysis are not shown here. They are very similar to those of the 
9-dimensional one but are less significant because the sample is reduced to 
one-half. We only comment that since the hardness has been added there, its 
weight has thus been enforced, and then class I becomes slightly harder and 
shorter than when obtained from 9 variables.

It must be noted that by just looking at the values of the dispersions in the
four variables T$_{\rm 90}$, H$_{\rm 32}$, P$_{\rm 1024}$ and F$_{\rm total}$
given in Table 2, it might seem that the variance would increase when shifting
from the two-class to the three-class classification, but that is just an
effect of projecting the groups onto these particular variables (three of
them composites): the full 9-dimensional analysis shows the opposite, as 
stated above. 

\begintable*{2}
\caption{{\bf Table 2.} Characteristics of the classification from the 
9-dimensional cluster analysis. T$_{{\rm 90}}$ is in units of s, 
P$_{{\rm 1024}}$ in photons cm$^{{\rm -} {\rm 2}} $s$^{{\rm -} {\rm 1}}$, and 
F$_{{\rm total}}$ in units of 10$^{{\rm -} {\rm 6}}$ erg cm$^{{\rm -2}}$.}
\halign{%
\rm#\hfil&\qquad\rm\hfil#&\qquad\rm\hfil#&\qquad\rm\hfil
#&\qquad\rm\hfil#&\qquad\rm\hfil#&\qquad\rm\hfil#
&\qquad\rm\hfil#&\qquad\rm\hfil#&\qquad\hfil#\rm\cr
\noalign{\hrule}
\noalign{\hrule}
\noalign{\vskip 2pt}
{\bf Class}&{\bf N}&{\bf $<$T$_{{\rm 90}}$$>$}&{\bf $<$H$_{{\rm 32}}$$>$}&{\bf $<$V/V$_{{\rm max}}$$>$}&{\bf $<$P$_{{\rm 1024}}$$>$}&{\bf $<$F$_{{\rm total}}$$>$}&{\bf $<$cos $\theta $$>$}&{\bf $<$sin$^{{\rm 2}{\rm} }$ b$-$1/3$>$}\cr
\noalign{\vskip 2pt}
\noalign{\hrule}
\noalign{\hrule}
\noalign{\vskip 3pt}
2$-$I&580&2.65$\pm $0.17&5.96$\pm $0.20&0.265$\pm $0.017&1.29$\pm $0.08&1.75$\pm $0.13&$-$0.031$\pm $0.026&$-$0.006$\pm $0.013\cr
2$-$II&1019&59.7$\pm $2.1&3.11$\pm $0.05&0.184$\pm $0.008&3.33$\pm $0.20&22.5$\pm $1.8&$-$0.004$\pm $0.019&+0.001$\pm $0.010\cr
\noalign{\vskip 2pt}
\noalign{\hrule}
\noalign{\vskip 2pt}
\noalign{\hrule}
\noalign{\vskip 2pt}
3$-$I&580&2.65$\pm $0.17&5.96$\pm $0.20&0.265$\pm $0.017&1.29$\pm 
$0.08&1.75$\pm $0.13&$-$0.031$\pm $0.026&$-$0.006$\pm $0.013\cr
3$-$II&570&51.3$\pm $2.3&2.85$\pm $0.07&0.296$\pm $0.012&0.88$\pm 
$0.02&4.58$\pm $0.21&+0.021$\pm $0.025&$-$0.000$\pm $0.013\cr
3$-$III&449&70.3$\pm $3.8&3.43$\pm $0.06&0.051$\pm $0.004&6.44$\pm 
$0.41&45.3$\pm $3.9&$-$0.035$\pm $0.030&+0.002$\pm $0.015\cr
\noalign{\vskip 3pt}
\noalign{\hrule}
\noalign{\hrule}
}
\endtable

Adopting the same expected values for isotropy as for the 4B catalog, that 
is for the Galactic dipole moment $<$cos $\theta $$>$ = $-$0.009 and for the 
quadrupole Galactic moment $<$sin$^{{\rm 2}{\rm} }$ b$-$1/3$>$ = $-$0.004, it 
can be seen in Table 2 that only one of the corresponding values for classes 
3I-3III lies beyond 1$\sigma$ of the expected value, and that is the dipole 
for the 3-II class, which is +1.2$\sigma$ above. Just such value being above 
1$\sigma$ appears not significant and it is concluded that all three classes 
are isotropically distributed.

In calculating the $<$V/V$_{{\rm m}{\rm a}{\rm x}}$$>$ parameter, not all 
the 1599 bursts could be used, but only those for which that value could 
be derived, and in a similar way when calculating the dipole and quadrupole 
moments the GRBs that were overwrites (Fishman et al. 1999) were not taken 
into account.

In Table 2 the class numbers are given in the order of increasing durations 
$<$T$_{{\rm 90}}$$>$. With the two-class classification the 'classical' GRB 
types are recovered: short/hard, which are fainter (taking as brightness 
the peak flux $<$P$_{{\rm 1024}}$$>$), and long/soft which are brighter and 
more non-Euclidean in their space distribution. As seen from Fig. 3, two 
classes with an overlapping distribution of durations have been obtained, in 
contrast with the classical definition of short, T$_{{\rm 9}{\rm 0}}$ $<$ 2 
s, and long, T$_{{\rm 9}{\rm 0}}$ $>$ 2 s, GRBs. Now the short class has 
durations up to $\sim$20 s while the long-duration class starts at $\sim$2 s. 
This overlapping of the two classes was obviously supposed to exist, but 
based on the distribution of durations alone it could not be decided whether,
in the overlapping region, a given GRB belonged to either of the two classes.
Now the algorithm handles all the available magnitudes and assigns each GRB 
to the cluster to whose characteristics it is closer to. The hardness 
distribution (Fig. 4) does not differ significantly from that in K93.

\beginfigure{3}

\input epsf
\epsfxsize=65mm \epsfbox{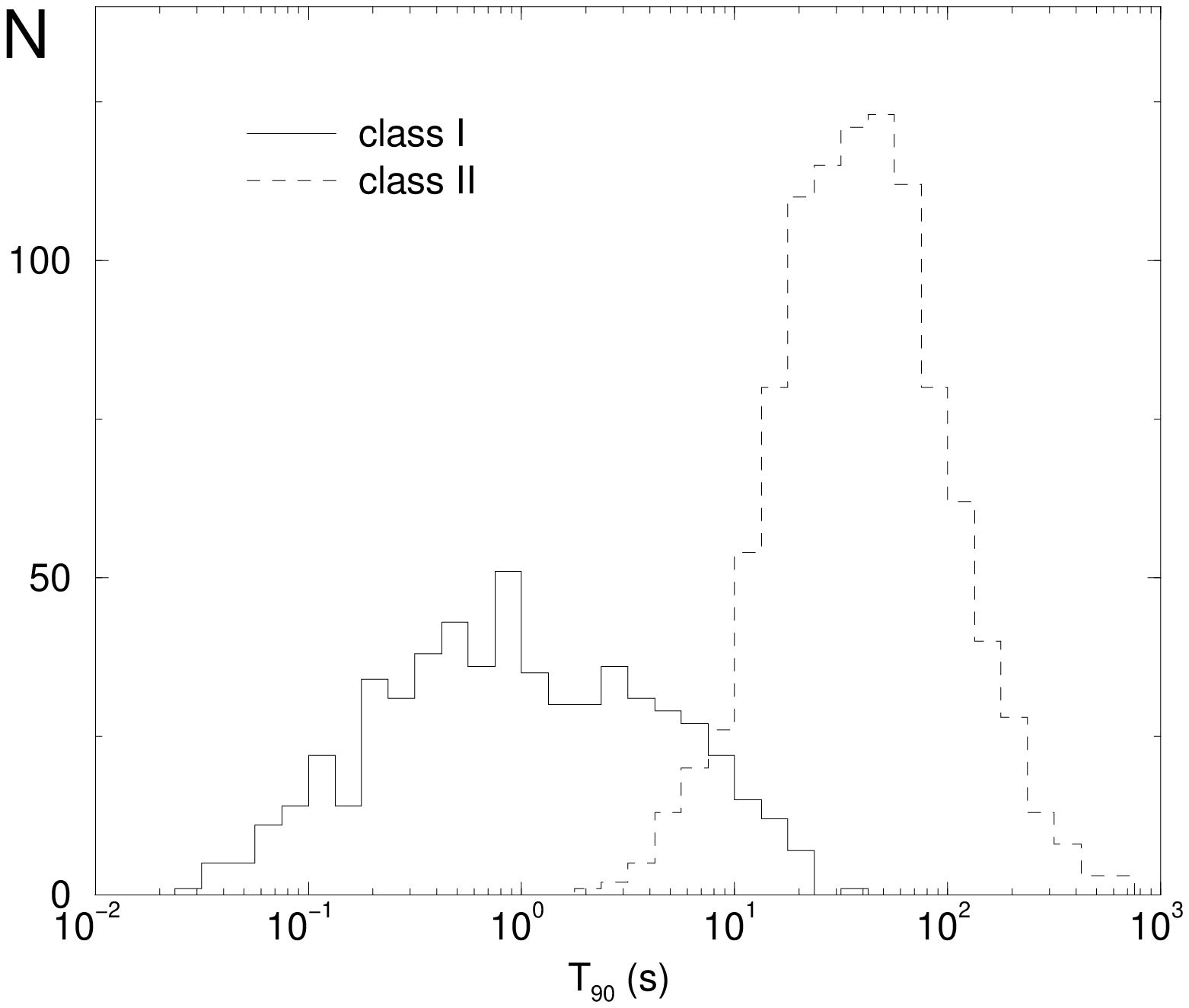}
\caption{{\bf Figure 3.} Duration distributions of classes 2-I and 2-II.}

\endfigure

\beginfigure{4}

\input epsf
\epsfxsize=65mm \epsfbox{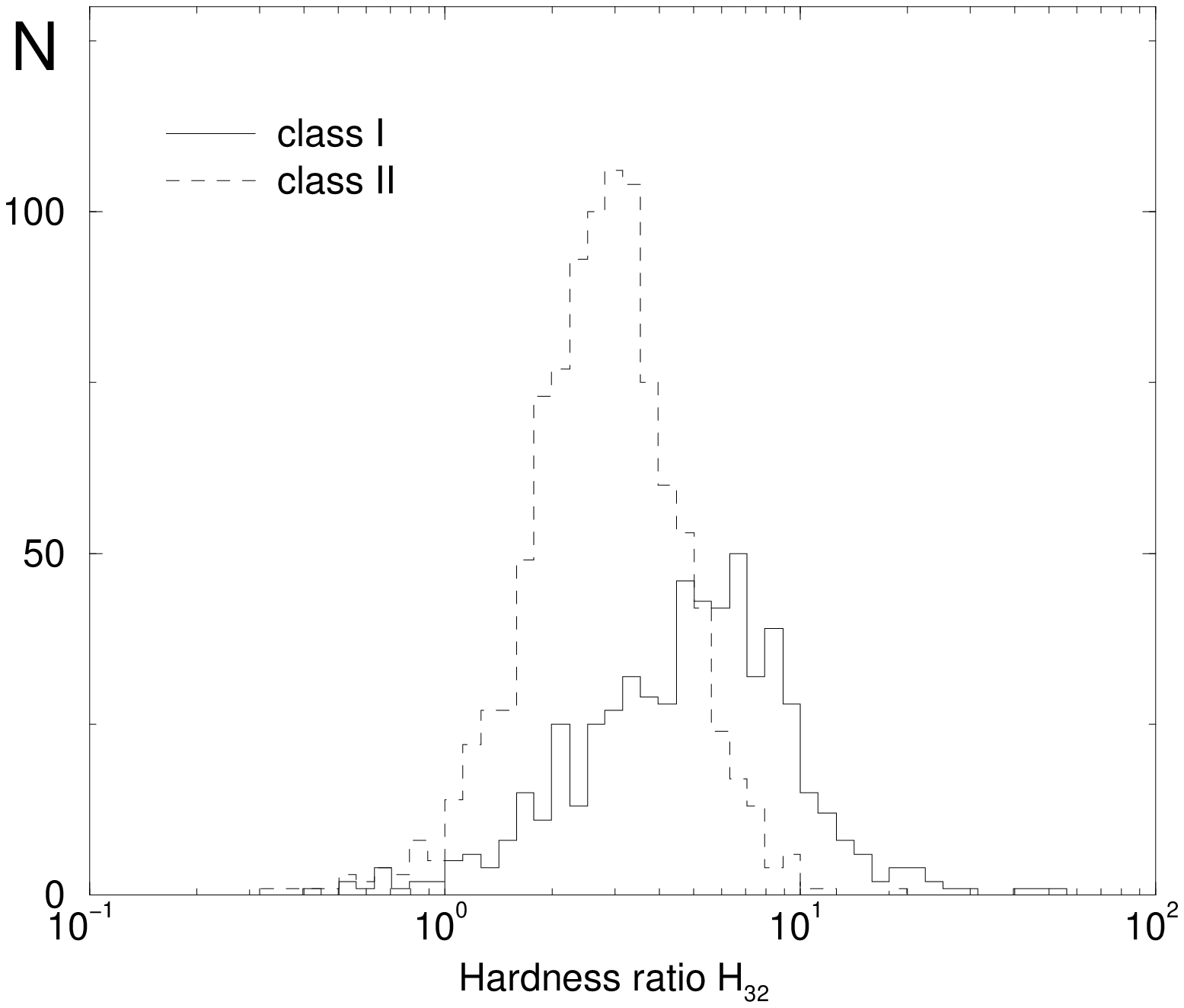}
\caption{{\bf Figure 4.} Hardness distributions of classes 2-I and 2-II.}

\endfigure

A first look at the three-class classification reveals that class I is 
exactly the same as in the two-class grouping: this is because the clustering 
method is agglomerative, which means that new groups are formed by merging 
former ones, so the passage from three to two classes happens when merging 
class II and class III GRBs.

Let us concentrate on the new three-class classification. As stated already, 
'old' class II has been divided into class II and class III. Class II is not 
properly an intermediate class: it has intermediate duration but still of the 
same order of magnitude as class III, and with an almost coincident 
distribution, as seen in Fig. 5. Class II is the softest and faintest class 
and the one most homogeneously distributed in space. Despite its duration 
being of the same order as that of class III, the fluence is one order of 
magnitude lower.

\beginfigure{5}

\input epsf
\epsfxsize=65mm \epsfbox{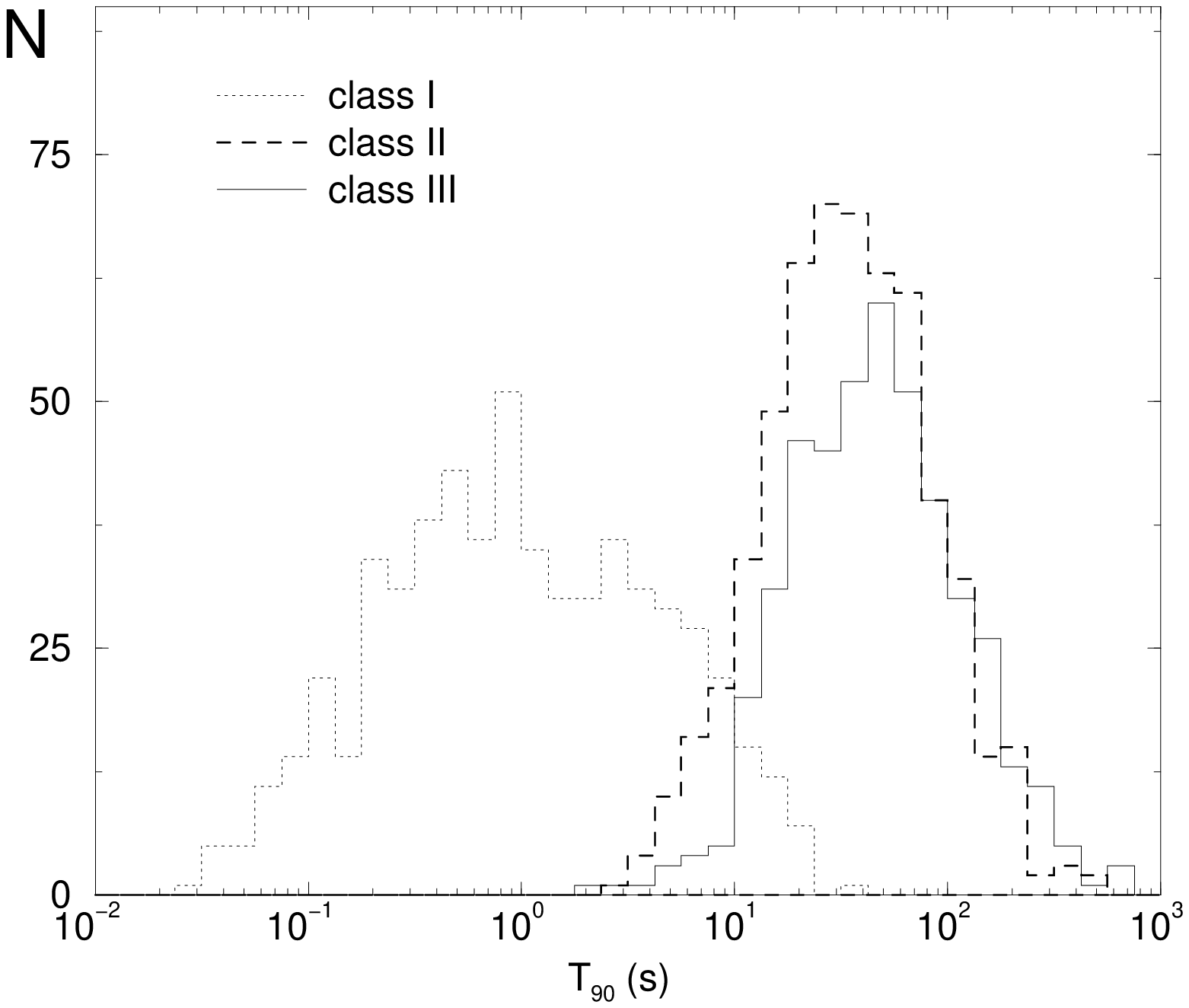}
\caption{{\bf Figure 5.} Duration distributions of the three-class 
classification from cluster analysis.}

\endfigure

\beginfigure{6}

\input epsf
\epsfxsize=65mm \epsfbox{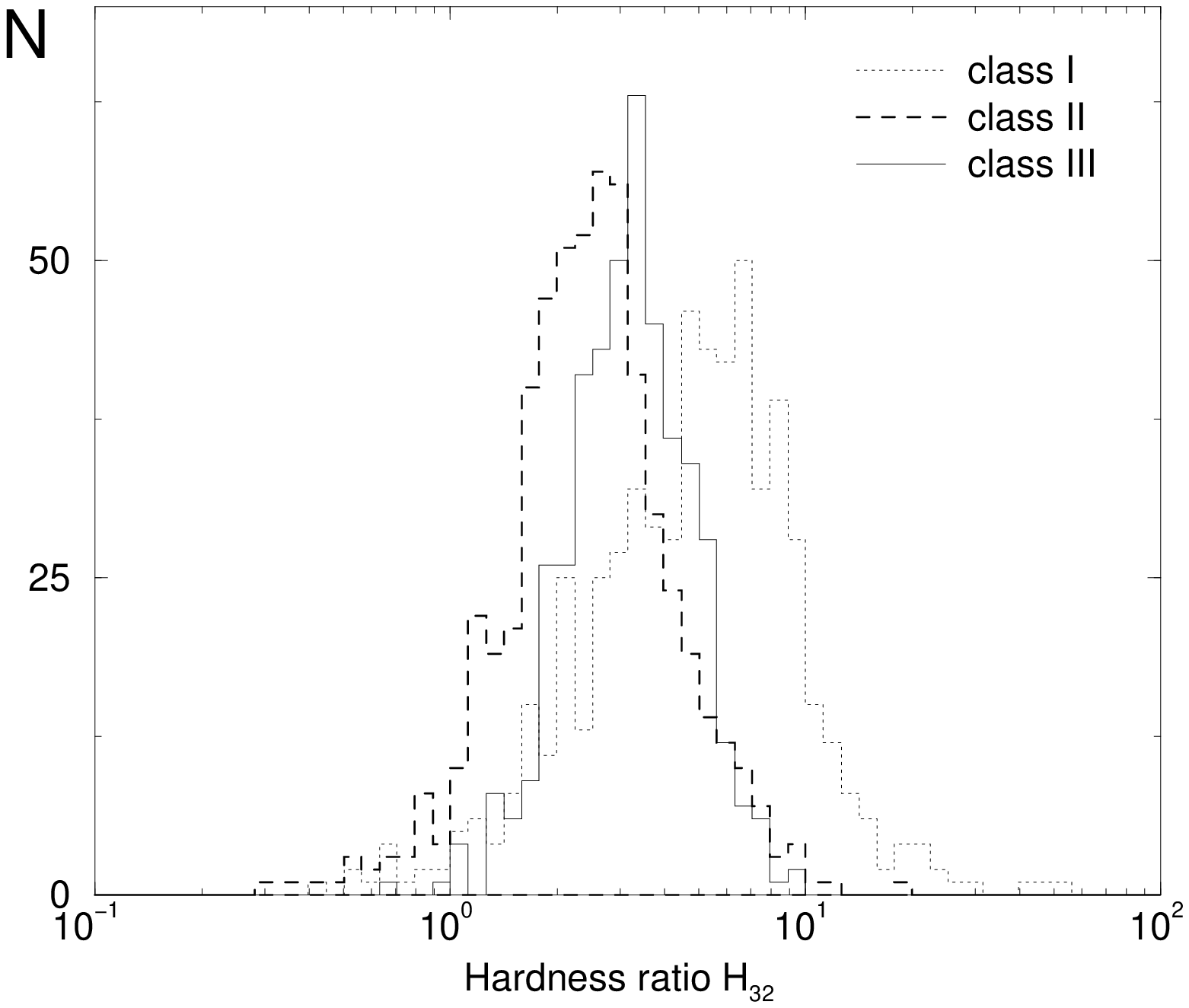}
\caption{{\bf Figure 6.} Hardness distributions of the three-class 
classification from cluster analysis.}

\endfigure

The most striking result of this new classification is the extremely low 
value of $<$V/V$_{{\rm m}{\rm a}{\rm x}}$$>$ in class III, which means that we 
are dealing with an extremely deep population that extends up to very high 
redshifts. Following the same procedure as in Mao \& Paczy{\'n}ski (1992), 
that is to calculate the theoretical value of 
$<$(F$_{{\rm min}}$/F)$^{{\rm 3/2}}$$>$ as a function of  
z$_{{\rm m}{\rm a}{\rm x}}$, and 
taking $<$V/V$_{{\rm m}{\rm a}{\rm x}}$$>$ in Table 2 as an empirical value 
for that quantity, a value of z$_{{\rm m}{\rm a}{\rm x}}$ for the 
distribution of the GRBs can be derived. Adopting  a model Universe with 
$\Omega _{{\rm M}}$ = 0.3, $\Omega _{{\rm \Lambda}}$ = 0.7, 
and H$_{{\rm 0}}$ = 65 Km s$^{{\rm -1}}$ Mpc$^{{\rm -1}}$, the GRBs to be
standard candles with a spectral slope $\alpha $ = 1 
(Mallozzi et al. 1996), and assuming constant comoving GRB rate, one obtains: 
z$_{{\rm m}{\rm a}{\rm x}}$ = 4.06$_{-0.57}^{+0.66}$ for class I, 
z$_{{\rm m}{\rm a}{\rm x}}$ = 3.08$_{-0.32}^{+0.35}$ for class II, and 
z$_{{\rm m}{\rm a}{\rm x}}$ = 45.24$_{-3.55}^{+4.23}$ for class III. The 
latter is an exceedingly high value, but as will be seen in \S 4, class III 
can have very massive stars as progenitors, and in that case the GRB rate 
should be proportional to the star formation rate (SFR) rather than being 
constant. Taking as SFR(z) that of Madau \& Pozzetti (2000), it is obtained 
for class III z$_{{\rm m}{\rm a}{\rm x}}$ = 11.30$_{-0.43}^{+0.56}$, which is 
a more reasonable value.

It was already known that separating long-class GRBs into two hardness groups 
results in two very different degrees of inhomogeneity (T98). The 
$<$V/V$_{{\rm m}{\rm a}{\rm x}}$$>$ values for the whole Current GRB Catalog 
have been calculated and the results are $<$V/V$_{{\rm m}{\rm a}{\rm x}}$$>$ = 
0.268$\pm $0.011, for GRBs with H$_{{\rm 3}{\rm 2}}$ $<$ 3, and 
$<$V/V$_{{\rm m}{\rm a}{\rm x}}$$>$ = 0.182$\pm $0.012, for GRBs with 
H$_{{\rm 3}{\rm 2}}$ $>$ 3. Now, with the three-class classification,  
very different degrees of inhomogeneity within the same interval of hardness 
are found, as it can be seen by comparing Table 2 with Fig. 6, and with little 
difference in mean hardness: those of class II and class III differ here with 
4$\cdot $10$^{{\rm -} {\rm 3}}$ significance in a Student test, but in the 
neural network classification the significance will only be of 0.42 while 
the $<$V/V$_{{\rm m}{\rm a}{\rm x}}$$>$ difference will remain. We leave a 
discussion of the value of $<$V/V$_{{\rm m}{\rm a}{\rm x}}$$>$ as related to 
hardness for \S 4.

It can also be pointed out that in contrast with what happened in the binary 
long/short classification, where shorter bursts were harder, now class II is 
shorter than class III but it is slightly softer at the same time.

\subsection{Neural Network}

In the neural network case, how many classes are to be obtained must be 
decided beforehand: knowing the dendrograms that result from the cluster 
analysis, we ask for either two or three classes. Their main characteristics 
are summarized in Table 3, using the same units as in Table 2.

\begintable*{3}
\caption{{\bf Table 3.} Characteristics of the classification with the neural 
network. T$_{{\rm 90}}$ is in units of s, P$_{{\rm 1024}}$ in photons 
cm$^{{\rm -} {\rm 2}}$ s$^{{\rm -} {\rm 1}}$, and F$_{{\rm total}}$ in units 
of 10$^{{\rm -} {\rm 6}}$ erg cm$^{{\rm -2}}$.}
\halign{%
\rm#\hfil&\qquad\rm\hfil#&\qquad\rm\hfil#&\qquad\rm\hfil
#&\qquad\rm\hfil#&\qquad\rm\hfil#&\qquad\rm\hfil#
&\qquad\rm\hfil#&\qquad\rm\hfil#&\qquad\hfil#\rm\cr
\noalign{\hrule}
\noalign{\hrule}
\noalign{\vskip 2pt}
{\bf Class}&{\bf N}&{\bf $<$T$_{{\rm 90}}$$>$}&{\bf $<$H$_{{\rm 32}}$$>$}&{\bf
$<$V/V$_{{\rm max}}$$>$}&{\bf $<$P$_{{\rm 1024}}$$>$}&{\bf 
$<$F$_{{\rm total}}$$>$}&{\bf $<$cos $\theta $$>$}&{\bf $<$sin$^{{\rm 2}{\rm}
}$ b$-$1/3$>$}\cr
\noalign{\vskip 2pt}
\noalign{\hrule}
\noalign{\hrule}
\noalign{\vskip 3pt}
2$-$I&685&6.24$\pm $0.50&5.50$\pm $0.18&0.288$\pm $0.015&0.94$\pm 
$0.04&1.44$\pm $0.09&+0.002$\pm $0.024&$-$0.005$\pm $0.012\cr
2$-$II&914&63.5$\pm $2.3&3.12$\pm $0.05&0.159$\pm $0.008&3.82$\pm 
$0.22&25.1$\pm $2.0&$-$0.024$\pm $0.021&+0.001$\pm $0.010\cr
\noalign{\vskip 2pt}
\noalign{\hrule}
\noalign{\vskip 2pt}
\noalign{\hrule}
\noalign{\vskip 2pt}
3$-$I&531&3.05$\pm $0.34&6.20$\pm $0.22&0.287$\pm $0.017&0.81$\pm 
$0.04&1.13$\pm $0.07&$-$0.003$\pm $0.027&$-$0.014$\pm $0.014\cr
3$-$II&341&25.0$\pm $1.4&3.05$\pm $0.10&0.307$\pm $0.019&1.25$\pm 
$0.08&2.82$\pm $0.16&$-$0.012$\pm $0.033&+0.009$\pm $0.016\cr
3$-$III&727&71.8$\pm $2.8&3.15$\pm $0.05&0.123$\pm $0.008&4.51$\pm 
$0.28&30.8$\pm $2.5&$-$0.022$\pm $0.023&+0.003$\pm $0.012\cr
\noalign{\vskip 3pt}
\noalign{\hrule}
\noalign{\hrule}
}
\endtable

As it should be expected, there are some differences in the composition of 
the classes as compared with those obtained from the clustering method, since
the neural network method is not agglomerative. So, for instance, class I is 
no longer identical, in the two-group classification, to class I in the 
three-group scheme. Also, the 'short' GRBs which make up this class now have 
longer average durations than in the cluster analysis.

There is also some change from the results of the cluster analysis in the 
three-group classification. Classes II and III now become more widely 
separated in duration, basically due to the decrease in duration of class II. 
The difference in hardness between class II and class III, in contrast, has 
decreased.

As in the cluster analysis, all three classes are highly isotropic, with no 
value of the moments above 0.8$\sigma $ of the values expected for isotropy.

\beginfigure{7}

\input epsf
\epsfxsize=65mm \epsfbox{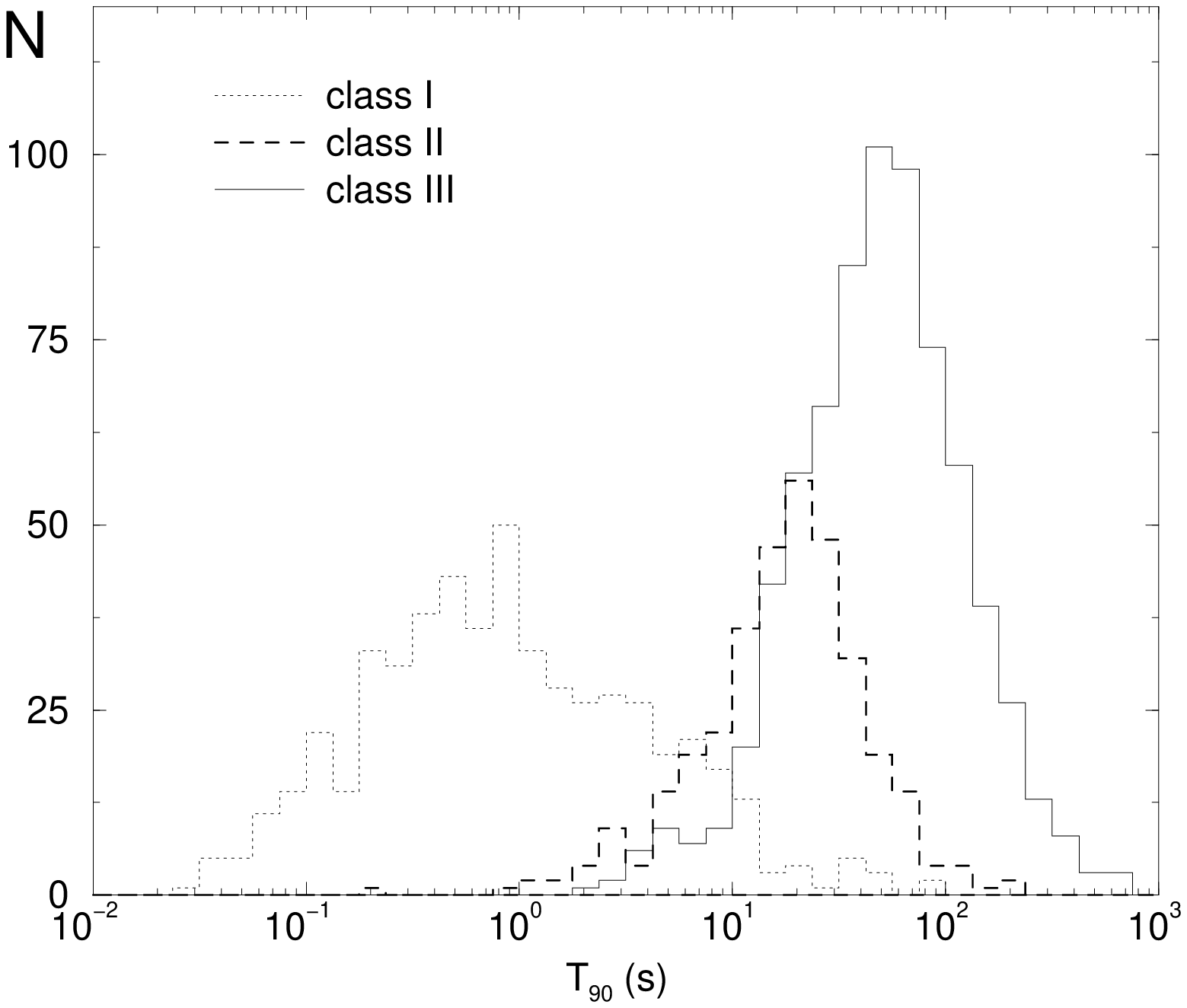}
\caption{{\bf Figure 7.} Duration distributions of the three-class 
classification from neural network analysis.}

\endfigure

\beginfigure{8}

\input epsf
\epsfxsize=65mm \epsfbox{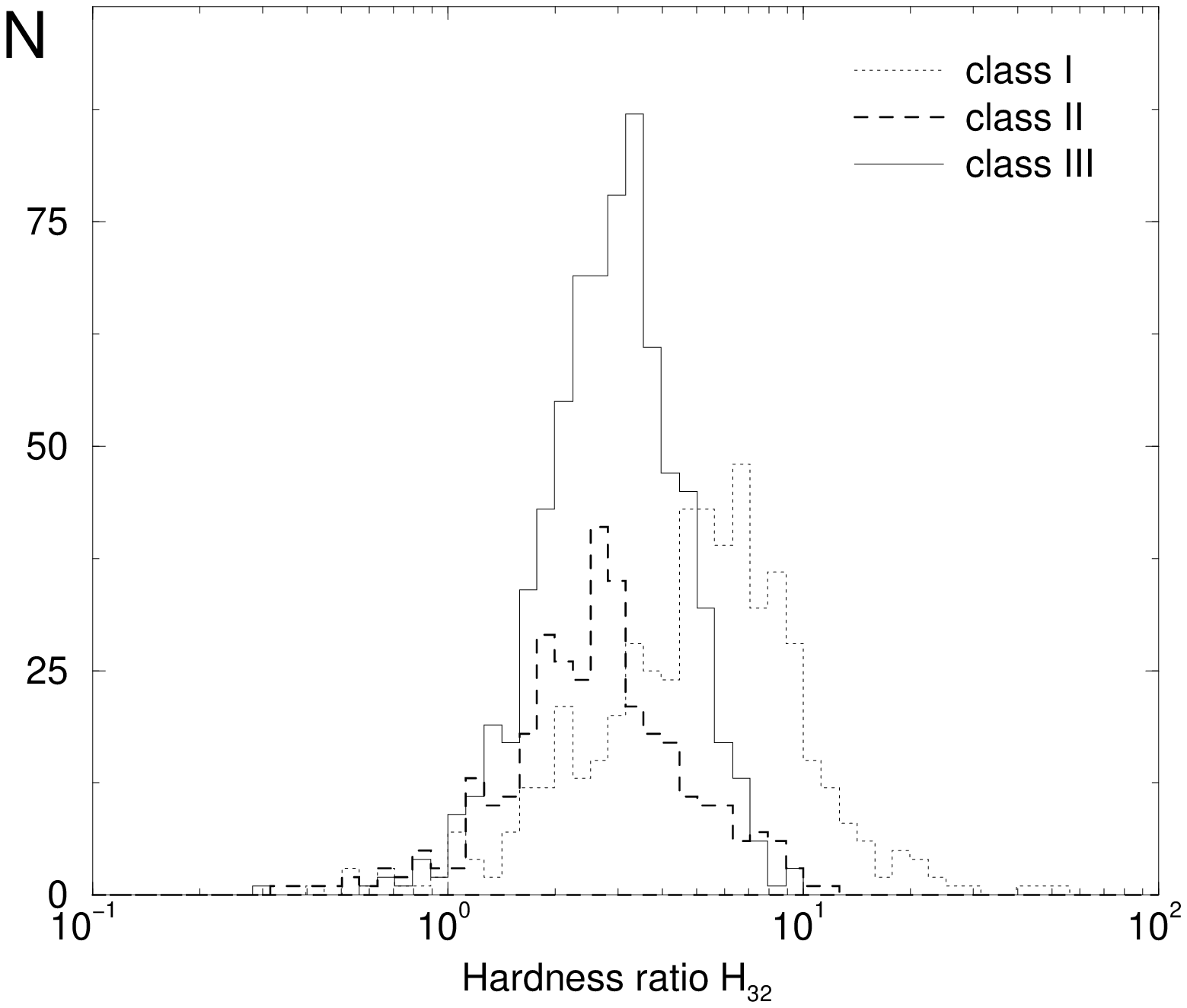}
\caption{{\bf Figure 8.} Hardness distributions of the three-class 
classification from neural network analysis.}

\endfigure

Class II is now the intermediate class in peak flux. From both methods, 
cluster analysis and neural network, it is seen that despite the difference 
by one order of magnitude between durations of class I and class II, their 
respective total fluences remain of the same order. 

The high inhomogeneity in the space distribution of class III is seen once 
more, and also how class II is again the most homogeneous one. Now the sample 
depths are: for constant comoving GRB rate, z$_{{\rm m}{\rm a}{\rm x}}$ = 
3.34$_{-0.48}^{+0.55}$, 2.78$_{-0.46}^{+0.53}$, and 15.45$_{-1.33}^{+1.50}$ 
for classes I, II, and III respectively. For GRB rate proportional to the SFR, 
z$_{{\rm m}{\rm a}{\rm x}}$ = 3.22$_{-0.20}^{+0.22}$, 2.98$_{-0.20}^{+0.23}$, 
and 6.67$_{-0.30}^{+0.30}$. The value of 
z$_{{\rm m}{\rm a}{\rm x}}\sim$ 11.3, obtained from the cluster analysis, 
corresponds to a Universe with an age of 4.3$\cdot $10$^{{\rm 8}}$ yr, and 
is in good agreement with the expectation of GRBs occuring out to at least 
z $ \approx $ 10. The value obtained from the neural network analysis of 
z$_{{\rm m}{\rm a}{\rm x}} \sim$ 6.7 corresponds to a Universe with an age 
of 8.8$\cdot $10$^{{\rm 8}}$ yr. Both values for z$_{{\rm m}{\rm a}{\rm x}}$ 
are below the redshift limit, z $\approx$ 15-20, given for Population III 
stars by LR00.

We conclude that the three classes respectively obtained from the cluster 
analysis and from the neural network algorithm show similar characteristics 
and thus both treatments are mutually consistent.

\section{Discussion}

The two different automatic classifier methods above suggest the existence of 
three different groups of GRBs with different properties. It will now be 
examined whether such a classification does make physical sense and we are 
actually dealing with three classes of GRBs.

Let us first discuss the proposal by T98, of taking the long-burst 
class and dividing it into two groups with H$_{{\rm 3}{\rm 2}}$ higher and 
lower than 3 respectively. As seen in \S 3.1, long/hard bursts are more 
inhomogeneously distributed than long/soft bursts. This might seem to be in 
contradiction with the cosmological scenario, in which more distant bursts 
are expected to be softer due to the spectrum redshift, and it leads us to 
conclude that class III bursts are intrinsically much harder than those of 
classes I and II.

Fig. 9 shows that there is indeed evolution in hardness: the value of 
$<$V/V$_{{\rm m}{\rm a}{\rm x}}$$>$ decreases with increasing 
H$_{{\rm 3}{\rm 2}}$. The hardness bins are taken so as to include similar 
numbers of bursts ($\sim$60) in each of them, in order to have comparable 
error bars. The value of $<$V/V$_{{\rm m}{\rm a}{\rm x}}$$>$ is displayed in 
the position of the mean of the hardness for each bin, and no error bars 
for the hardness are shown because the deviation is less than the symbol 
size, except for the last bin for which it is of about 0.2

\beginfigure{9}

\input epsf
\epsfxsize=65mm \epsfbox{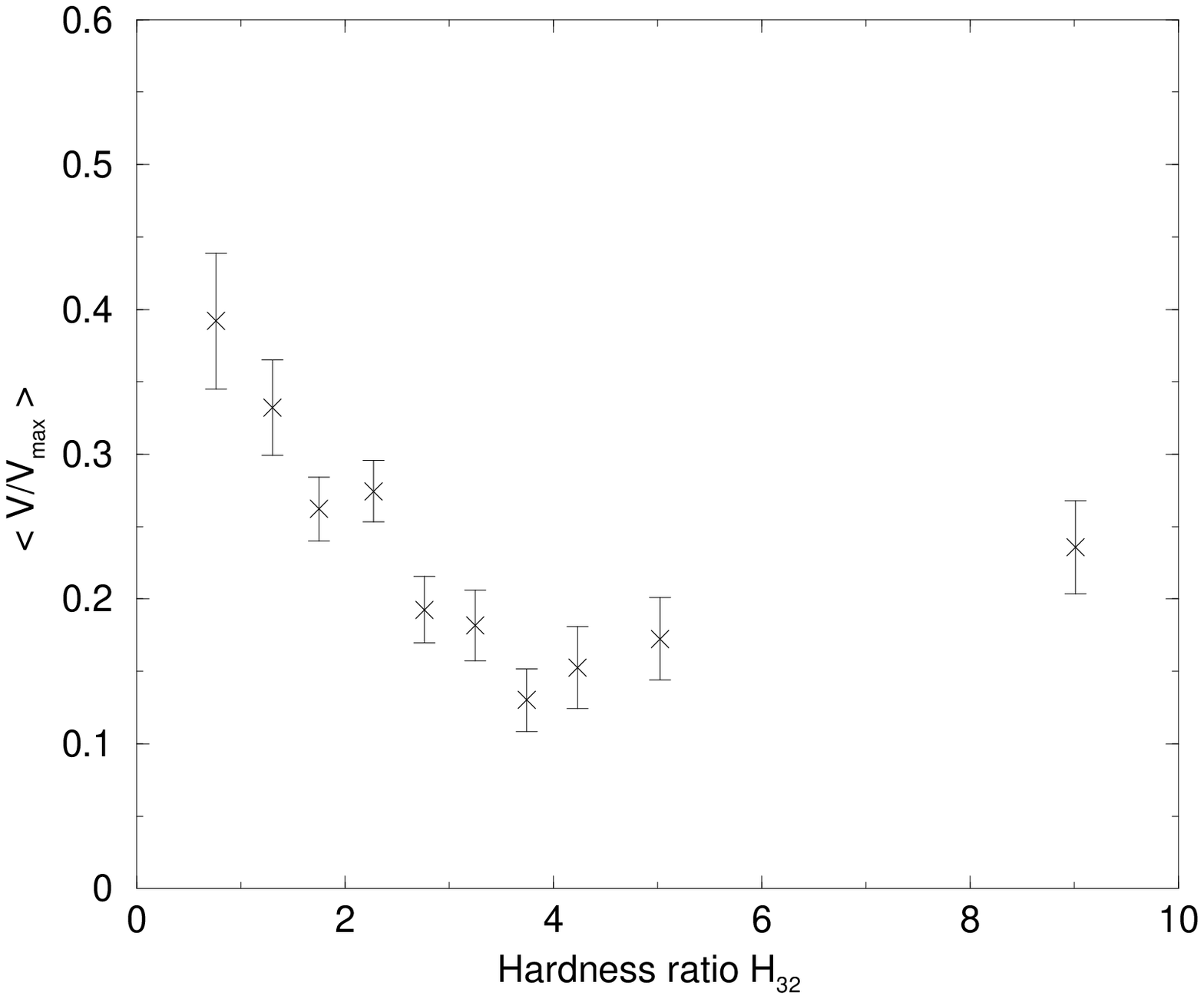}
\caption{{\bf Figure 9.} $<$V/V$_{{\rm m}{\rm a}{\rm x}}$$>$ vs. Hardness for 
GRBs with T$_{{\rm 9}{\rm 0}}$ $>$ 2 s. The correlation between these two 
variables is clearly seen. Hardness bins have been taken so as to include 
similar numbers of bursts in all of them. Each hardness bin contains 
$\sim$60 GRBs.}

\endfigure

When $<$V/V$_{{\rm m}{\rm a}{\rm x}}$$>$ decreases one is dealing 
with a more distant sample of GRBs, and then Fig. 9 tells us that, when
sampling to higher distances, GRBs tend to be harder, and taking into 
account the hardness-intensity correlation (Dezalay et al. 1997) they should 
also be more luminous. This effect has to be interpreted, in a cosmological 
scenario, as a source evolution. There is a possible explanation: it is  
generally admitted that the upper limit of the stellar initial mass function 
(IMF) depends on metallicity, and that lower metallicity allows more massive 
stars to form. When sampling GRBs farther away, one looks to a younger 
Universe, with lower metallicity, and thus with more massive stars. Therefore,
if GRBs come from very massive stars those ancient GRBs had sources with 
higher power and they were brighter and harder.

Next are displayed, in Fig. 10 and Fig. 11, 
$<$V/V$_{{\rm m}{\rm a}{\rm x}}$$>$ versus hardness for classes II and III, 
from the results of the cluster and of the neural network analyses, 
respectively.

\beginfigure{10}

\input epsf
\epsfxsize=65mm \epsfbox{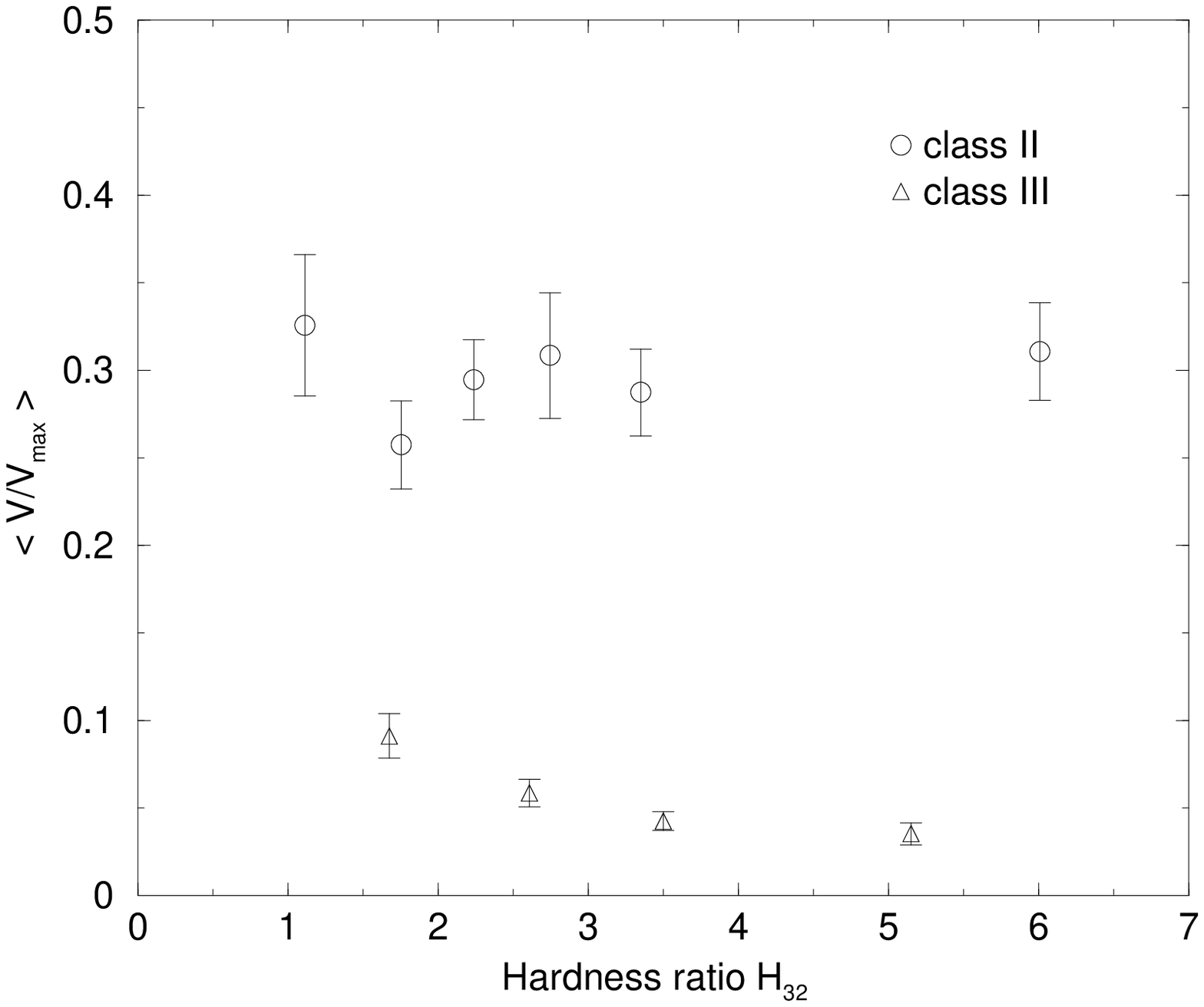}
\caption{{\bf Figure 10.} $<$V/V$_{{\rm m}{\rm a}{\rm x}}$$>$ vs. Hardness 
for classes II and III from the cluster analysis. Here the nearly 
constant value of $<$V/V$_{{\rm m}{\rm a}{\rm x}}$$>$ over the interval of 
hardnesses covered by class II is seen. The trend for class III of lower 
$<$V/V$_{{\rm m}{\rm a}{\rm x}}$$>$ with higher H$_{{\rm 3}{\rm 2}}$ can 
also be seen. As in Figure 9, the hardness bins have been taken so as to
include similar numbers of bursts in all of them.}

\endfigure

\beginfigure{11}

\input epsf
\epsfxsize=65mm \epsfbox{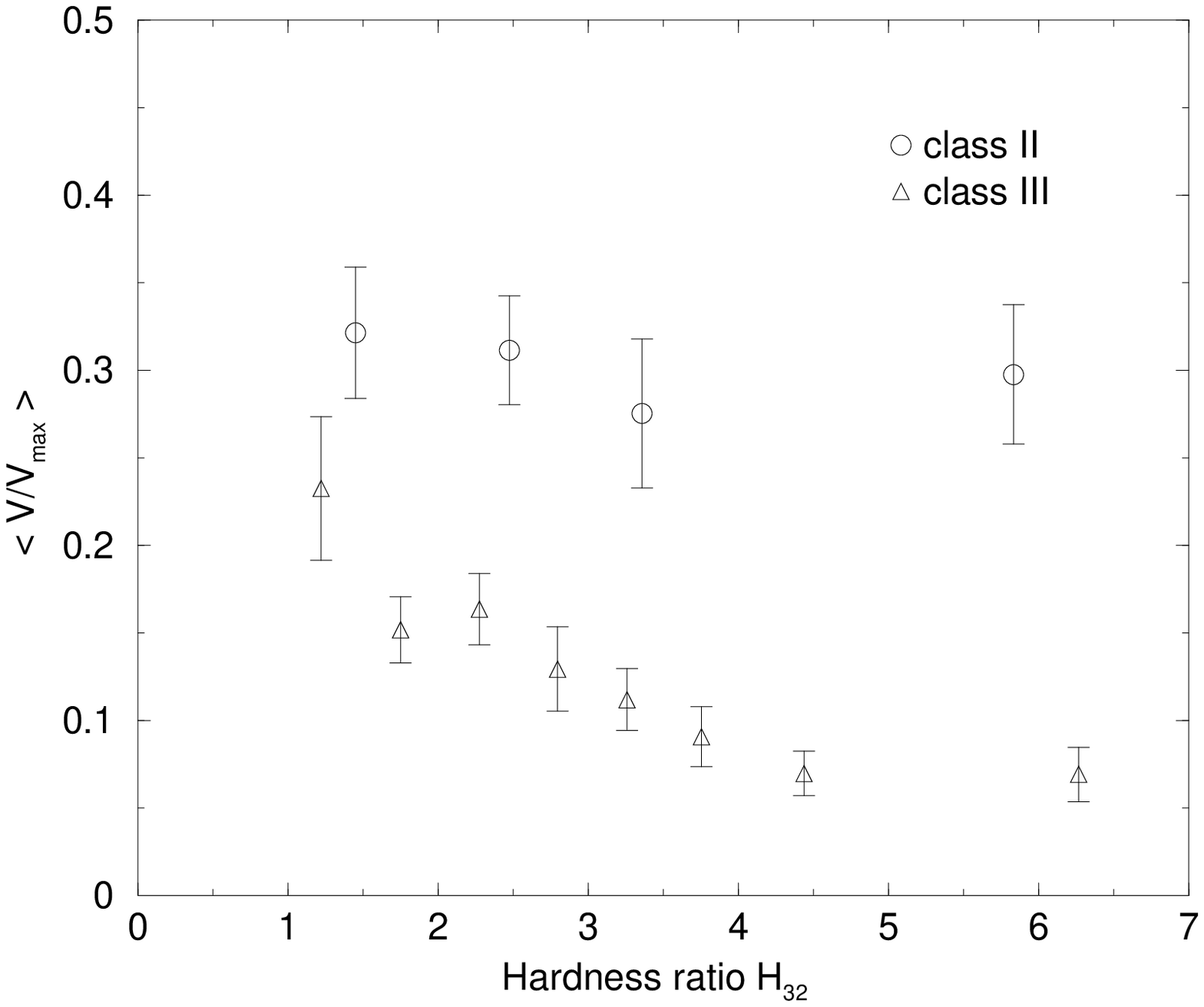}
\caption{{\bf Figure 11.} $<$V/V$_{{\rm m}{\rm a}{\rm x}}$$>$ vs. Hardness 
for classes II and III from neural network analysis. In this graph, the trend 
for class III of lower $<$V/V$_{{\rm m}{\rm a}{\rm x}}$$>$ with higher 
H$_{{\rm 3}{\rm 2}}$ is more evident than in Figure 10. Same criterion as
in Figure 10 for the hardness bins.}

\endfigure

Here can be appreciated, with particular clarity in the Figure corresponding 
to the neural network result, how in the three-class scheme the 'new' class 
II no longer shows any trend of $<$V/V$_{{\rm m}{\rm a}{\rm x}}$$>$ 
decreasing with increasing H$_{{\rm 3}{\rm 2}}$. Such a trend in the 'old' 
class II was due to the fusion into it of the 'new' classes II and III, and 
now it is seen that class III is the only one to uphold the trend. To evaluate 
numerically the correlation, a Spearman rank test (see, for instance, its 
implementation in Press et al. 1992) has been applied, obtaining for class 
III a Spearman-rank correlation coefficient r$_{{\rm s}}$ = $-$0.344, with a 
significance of 2$\cdot $10$^{{\rm -} {\rm 8}}$, for the class resulting from 
the cluster analysis, and r$_{{\rm s}}$ = $-$0.354, with significance 
4$\cdot $10$^{{\rm -} {\rm 1}{\rm 3}}$, for that resulting from the neural 
network analysis. In contrast, for class II from the cluster analysis 
r$_{{\rm s}}$ = 0.066 is obtained, with a significance of 0.26, and 
r$_{{\rm s}}$ = $-$0.051 with a 0.51 significance level for class II from the 
neural network. It is concluded, therefore, that class III really shows clues 
of cosmological source evolution, which can be due to its being made of GRBs 
produced by very massive stars, likely through collapsars.

Out of all bursts with known redshifts, eight of them entered into our 
classification. In the case of the neural network classification, seven out of 
those eight bursts are classified as class III, and only one of them as class 
II. With the clustering method classification, six belong to class III and 
two to class II. One of these last two bursts is GRB 980425, presumably 
related to SN 1998bw (Galama et al. 1999), which is thought to be a peculiar 
GRB. In both classifications GRB 970508 (Metzger et al. 1997) was assigned to 
class II. The assignation of any individual GRB to a given class by our 
methods is not entirely reliable, however, and has an uncertainty that is 
most important in the limiting region of each cluster. By looking at the 
scatter graphs, it has been checked that GRB 970508 lies in fact in the region 
near class III. It must be stressed that, given the high isotropy of all 
three classes found here, there is no clue of Galactic structure for any of 
them. It could be argued that, in the same way that no redshift has yet been 
measured for any GRB of class I, due to the fact that their being short makes 
them difficult to be detected with {\it BeppoSAX} (mainly sensitive to bursts 
longer than about 5-10 s), no redshifts of GRBs of class II have been 
measured either because they are faint and their detection is equally hard. 
An alternative explanation would be that the fact that no afterglows from 
GRBs of classes I or II have been seen is rather due to their being produced 
by NS-NS or NS-BH mergings, which are expected to happen mostly outside 
galaxies, where the interstellar medium is too tenuous to produce any 
detectable afterglow (see Panaitescu, Kumar \& Narayan 2001, for instance). 
That would also be consistent with the $<$V/V$_{{\rm m}{\rm a}{\rm x}}$$>$ 
values found: while collapsars should appear first and be more frequent in 
the distant, early Universe, NS-NS and NS-BH mergings should start later and 
be more homogeneously distributed down to low redshifts. In this context we 
can even speculate whether the differences between class I and class II GRBs 
might be due to one of them corresponding to BH-NS (or BH-WD) mergings, the 
other class being produced by NS-NS mergings. The whole question should be 
solved with the new generation of GRB detectors aboard the HETE II and {\it 
Swift} satellites.

One should be concerned whether the structure of the GRB data may partially
reflect instrumental biases. Hakkila et al. (2000) have suggested that the 
three-class classification obtained by M98 might arise from a bias in 
measuring some bursts properties, such as duration and fluence, which would 
make some bursts in 'classical' class II to take the 'new' class II 
characteristics (by lowering their duration and fluence). The 
fluence-duration bias, however, in spite of being qualitatively understood, 
is not well quantified. Hakkila et al. based their analysis on M98 classes:  
their intermediate duration class has durations T$_{{\rm 9}{\rm 0}}$ of 
between 2 s and 10 s, while the class II deduced here extends up to 
T$_{{\rm 9}{\rm 0}}$ longer than 100 s. Moreover, such bias acts on the 
farthest bursts, while what we find is that our 'new' class II GRBs are the 
closest ones. In addition, any bias that would make some bursts in the 'old' 
class II appear shorter and with lower fluence could hardly separate 
at the same time the evolutionary effects that we see in Fig. 9 into two 
groups: one with evolution (class III) and the other one without it ('new' 
class II).  

\section{Conclusion}

There are reasons to think that there exists more than one type of possible 
progenitors for GRBs, and each type may give rise to groups of burst with 
different properties. We have searched for those groups in the current BATSE 
catalog, with the aid of two automated classification algorithms, and  
confirmed that there exist two clearly separated classes of GRBs corresponding 
to the 'classical' classification of long/short GRBs. In addition, we have 
also obtained clear hints that there exists a third class, different from 
those previously reported. An oversimplified way of looking at this would be
to say that the third class arises from splitting the original long class 
into two groups with high and low peak fluxes respectively, in a similar way 
that the whole sample of GRBs has been divided, in previous studies, into 
pairs of groups according to duration (K93), hardness (T98), brightness 
(Nemiroff et al. 1994), or other characteristics (Pendleton et al. 1997). The 
present work, however, goes beyond that since nine quantities related to the 
bursts are used for the classification instead of taking a single parameter 
and then finding a value separating the bursts into two classes: there are 
overlapping zones in every original variable. What our procedures do is to 
trace a surface in the 9-dimensional space, separating classes from the way 
each variable relates to all others. Two different classes may well have the 
same duration or show nearly the same distribution for a given variable, but 
by taking into account the other variables as well these procedures still 
detect their existence. In contrast, univariate distributions would overlook 
them.

Apart from the power of the method, the new grouping of the bursts thus 
obtained has to be examined for its possible physical meaning and its 
correspondence with separate classes of GRB progenitors and/or mechanisms. 
Classes I, II, and III here defined correspond to different observational 
depths (z$_{{\rm m}{\rm a}{\rm x}}$) and may result from varying geometries 
of the observer with respect to the emitter, different parameters of the 
explosion, or from different progenitors having different spatial  
distributions. Thus, every class has to be compared with several possible 
models. The physical separation of classes II and III is strongly supported 
by the fact, which can hardly be due to chance alone, that having both classes 
together they show evolution of hardness and intensity with the maximum 
distance sampled while when separated such evolution only exists in class 
III. We conclude, therefore, that class III, which likely has collapsars as 
progenitors, is the one that can be detected up to very large redshifts, and 
it should thus be the most adequate one to learn about the history of the 
Universe at high z. We also suggest that classes I and II could correspond to 
NS-NS or NS-BH mergings instead.

\section*{Acknowledgements}

We thank J. Bloom for suggestions and comments. Andreu Balastegui thanks R. 
Cabez\'{o}n for helpful discussions and comments.


\section*{References}

\beginrefs

\bibitem Bagoly Z., M\'{e}sz\'{a}ros A., Horv\'{a}th I., Bal\'{a}zs L. G., 
M\'{e}sz\'{a}ros P., 1998, ApJ, 498, 342 (B98)

\bibitem Berezinsky V., Hnatyk B., Vilenkin A., 2001, (astro-ph/0102366)

\bibitem Blinnikov S., 1999, (astro-ph/9902305)

\bibitem Bloom J. S., Kulkarni S. R., Djorgovski S. G., 2000, 
(astro-ph/0010176)

\bibitem Castro-Tirado, A., 2001, in Exploring the Gamma-ray Universe, in 
press, (astro-ph/0102122)

\bibitem Daigne, F., Mochkovitch, R., 1998, MNRAS, 296, 275 

\bibitem Dezalay J. P. et al., 1997, ApJ, 490, L17

\bibitem Fishman G. J. et al., 1999, ApJSS, 122, 465

\bibitem Galama T. J., et al., 1999, A\&ASS, 138, 465

\bibitem Hakkila J., Haglin D. J., Pendleton G. N., Mallozzi R. S., Meegan C. 
A., Roiger R. J., 2000, ApJ, 538, 165

\bibitem Horv\'{a}th I., 1998, ApJ, 508, 757

\bibitem Katz J. I., Canel L. M., 1996, ApJ, 471, 915

\bibitem Klebesadel R. W., Strong I. B.,Olson R. A., 1973, ApJ, 182, L85

\bibitem Kohonen T., 1990, IEEC Proceedings, 78, 1464

\bibitem Kouveliotou C., Meegan C. A., Fishman G. J., Bhat N. P., Briggs M. 
S., Koshut T. M., Paciesas W. S., Pendleton G. N., 1993, ApJ, 413, L101 (K93)

\bibitem Lamb D. Q., Reichart D. E., 2000, ApJ, 536, 1 (LR00)

\bibitem Ma F., Xie B., 1996, ApJ, 462, L63

\bibitem MacFadyen A. I., Woosley S. E., 1999, ApJ, 524, 262

\bibitem Madau P., Pozzetti L., 2000, MNRAS, 312, L9

\bibitem Mallozzi R. S., Pendleton G. N., Paciesas W. S., 1996, ApJ, 471, 636

\bibitem Mao S., Paczy{\'n}ski B., 1992, ApJ, 388, L45

\bibitem Meegan C. A., Fishman G. J., Wilson R. B., 1985, ApJ, 291, 479

\bibitem M\'{e}sz\'{a}ros P., 2001, Science, 291, 79 

\bibitem M\'{e}sz\'{a}ros P., Rees, M. J., 1993, ApJ, 405, 278

\bibitem Metzger M. R. et al. 1997, IAU Circ. 6655

\bibitem Mukherjee S., Feigelson E. D., Babu G. J., Murtagh F., Fraley C., 
Raftery A., 1998, ApJ, 508, 314 (M98)

\bibitem Murtagh F., Heck A., 1987, Multivariate Data Analysis, Astrophysics 
and Space Science Library, Dordrecht Reidel Publ. (M87)

\bibitem Narayan R., Paczy{\'n}ski B., Piran T., 1992, ApJ, L83

\bibitem Nemiroff R. J., 1994, (astro-ph/9402012)

\bibitem Nemiroff R. J., Norris J. P., Bonnell J. T., Wickramasinghe W. A. D. 
T., Kouveliotou C., Paciesas W. S., Fishman G. J., Meegan C. A., 1994, ApJ, 
435, L133

\bibitem Paczy{\'n}ski B., 1990, ApJ, 348, 485

\bibitem Paczy{\'n}ski B., 1990, ApJ, 363, 218

\bibitem Paczy{\'n}ski B., 1998, in Gamma Ray Bursts: 4th Huntsville 
Symposium, ed. C.A. Meegan, R.D. Preece and T.M. Koshut, AIP, New York, 783

\bibitem Panaitescu, A., Kumar, P., Narayan, R., 2001, (astro-ph/0108132) 

\bibitem Pendleton G. N. et al., 1997, ApJ, 489, 175

\bibitem Piran, T., 2000, Phys. Rep., 333, 529 

\bibitem Press W. H. et al., 1992, Numerical Recipes in Fortran, Second 
Edition, Cambridge University Press, Cambridge

\bibitem Rees M. J., M\'{e}sz\'{a}ros P., 1992, MNRAS, 258, 41{\sc p}

\bibitem Spruit, H.C., 1999, A\&A, 341, L1 

\bibitem Tavani M., 1998, ApJ, 497, L21 (T98)

\bibitem Totani T., 1997, ApJ, 486, L71

\bibitem Totani T., 1999, ApJ, 511, 41

\bibitem Ward, J.H., 1963, in Statistical Challenges of Modern Astronomy, ed.
G.J. Babu and E.D. Feigelson, Springer, New York, 135

\bibitem Woosley S. E., 1993, ApJ, 405, 273

\endrefs
\bye